# Picoscale control of quantum plasmonic photoluminescence enhancement at 2D lateral heterojunction

Zachary H. Withers,[a] Sharad Ambardar,[ab] Xiaoyi Lai,[d] Jiru Liu,[c] Alina Zhukova,[a] and Dmitri V. Voronine[ab]*

[a]Department of Physics, University of South Florida, Tampa, FL 33620, USA.

[b]Department of Medical Engineering, University of South Florida, Tampa, FL 33620, USA

[c]Department of Physics, Texas A&M University, College Station, TX 77843, USA

[d]School of the Gifted Young, University of Science and Technology of China, Hefei 230026, China

## Abstract

Two-dimensional (2D) materials and heterostructures have recently gained wide attention due to potential applications in optoelectronic devices. However, the optical properties of the heterojunction have not been properly characterized due to the limited spatial resolution, requiring nano-optical characterization beyond the diffraction limit. Here, we investigate the lateral monolayer $MoS_2$-$WS_2$ heterostructure using tip-enhanced photoluminescence (TEPL) spectroscopy on a non-metallic substrate with picoscale tip-sample distance control. By placing a plasmonic Au-coated Ag tip at the heterojunction, we observed more than three orders of magnitude photoluminescence (PL) enhancement due to the classical near-field mechanism and charge transfer across the junction. The picoscale precision of the distance-dependent TEPL measurements allowed for investigating the classical and quantum tunneling regimes above and below the ~320 pm tip-sample distance, respectively. Quantum plasmonic effects usually limit the maximum signal enhancement due to the near-field depletion at the tip. We demonstrate a more complex behavior at the 2D lateral heterojunction, where hot electron tunneling leads to the quenching of the PL of $MoS_2$, while simultaneously increasing the PL of $WS_2$. Our simulations show agreement with the experiments, revealing the range of parameters and enhancement factors corresponding to various regimes. The controllable photoresponse of the lateral junction can be used in novel nanodevices.

# Introduction

Lateral 2D heterostructures have recently found interest due to potential optoelectronic applications and their exceptional properties, which include atomically sharp junctions, quantum confinement and band gap tunability[1–13]. However, because of the highly averaged measurements of the far-field (FF) optical characterization experiments such as the conventional photoluminescence (PL) spectroscopy, there is a need to improve the understanding and applications of the heterojunctions at the nanoscale s. This is experimentally challenging since the PL signals from the atomically thin junctions are weak and the surrounding materials generate large background. Previously, several nano-optical imaging techniques were used to address this challenge including scanning near-field optical microscopy (SNOM)[14,15], tip-enhanced raman spectroscopy (TERS) [16–18]and tip-enhanced photoluminescence (TEPL)[19–21]. The spatial imaging resolution in these techniques depends on the size of the excitation spot and on the signal enhancement factor (EF). The excitation spot size is limited by the size of scanning local probe such as the plasmonic metallic tip in TEPL, which is typically on the order of ~10 nm. The signal enhancement is limited by the near-field (NF) enhancement, which depends on tip-sample distance. Classically, the PL signal enhancement is inversely proportional to the tip-sample distance[22]. However, for the tip-sample distances shorter than ~1 nm, the PL enhancement decreases due to charge tunneling between the tip and the sample, described using the quantum plasmonics model [23–32]. Additionally, hot electrons, generated in plasmonic systems, contribute to the PL enhancement[33–35].

Hot electron injection (HEI) in 2D materials has gained a wide interest due to the the applications of HEI in 2D semiconductors, especially transition metal dichalcogenides (TMDs), in photoemission of electrons and photosensitive reactions[36–38]. The strong localized electromagnetic fields of the plasmonic metal tips can be used for HEI into TMDs [39–41]. The HEI rate can be adjusted by varying the tip-sample distance in the nanoscale classical and picoscale quantum regimes.

Previously, we reported the HEI in lateral $MoSe_2$-$WSe_2$ heterostructures showing $MoSe_2$ PL enhancement and $WSe_2$ PL quenching in the quantum regime. However, the enhancement mechanism was not thoroughly investigated, and no EF limiting values were reported. Also, a different chemical vapor deposition (CVD) growth procedure was used leading to different properties of the junction and the lateral spatial dependence of the PL enhancement across the heterojunction was not studied. The $MoS_2$-$WS_2$ and $MoSe_2$-$WSe_2$ heterostructures have different electronic structures and therefore different resonant couplings with the plasmonic resonant optical antenna tips.

In this work, we performed the picometer-scale controlled tip-sample distance dependent TEPL measurements using a 660 nm laser excitation, to investigate the effect of tunneling hot electrons from the Au-coated plasmonic Ag tip to the CVD-grown $MoS_2$-$WS_2$ heterostructure on a non-conductive $SiO_2$/Si substrate. The experiments were carried out at several spatial locations across the heterojunction in the classical (360 pm< d < 20 nm) and quantum tunneling (220 pm< d < 360 pm) regimes. We calculated the EF distance dependence transition from the classical to the quantum regime for different HEI rates, which showed a good agreement with the experiments. We proposed a new charge transfer channel from $MoS_2$ to $WS_2$ to explain the observed PL distance dependence. This work provides new insights into the PL enhancement mechanisms at the 2D lateral heterojunctions and can be used for tuning 2D nanodevices.

## Materials and methods

Monolayer lateral $MoS_2$-$WS_2$ heterostructures were grown on $SiO_2$/Si substrates in a quartz tube using a one-pot CVD method as previously described[2].

The TEPL experiments were carried out using a commercial system (OmegaScope-R coupled with LabRam Evolution; Horiba Scientific). AFM measurements were carried out using Si tips with the tip apex radius of ~10 nm. The TEPL measurements were performed using Au-coated Ag tips with the tip apex radius of ~25 nm, coupled with 660 nm linearly polarized laser excitation beam, focused on the tip apex at the 53⁰ angle of incidence [29].

## Results and discussion

Figure 1a illustrates the picoscale controlled tip-sample distance dependent TEPL measurements on a monolayer lateral $MoS_2$-$WS_2$ heterostructure using three different tip locations on the $MoS_2$ (left), $WS_2$ (right) and center of the heterojunction (middle). The measured PL signals with the tip-sample distance, 0.36 nm < d < 20 nm, are referred to as the classical regime. As the metallic tip reaches the quantum regime, there is a charge transfer from $WS_2$ to $MoS_2$ across the heterojunction which is represented by the purple arrow ($\gamma_1 \Gamma_p(d)$) and the corresponding inverse process shown by the blue arrow, $\gamma_2 \Gamma_p(d)$. The hot electron injection with the rate of $G_{HEI}\Gamma_{CT}(d)$ occurs from the plasmonic tip the semiconductor sample, relaxing to excitons at rates α or β in $MoS_2$ and $WS_2$, respectively, or through nonradiative decay channels at the rate of $R_{HEI}$. The exciton transfer from $WS_2$ to $MoS_2$ at the heterojunction is assumed to be proportional to the near field optical excitation, $\Gamma_p(d)$. The tip-sample measurements were performed on 7 spots, with 4 spots on $WS_2$, 2 spots on $MoS_2$, and 1 spot (Spot 3) performed on the heterojunction. For the data analysis, we chose spots from 1 to 5, since spots 6 and 7 showed the same behavior as spot 1. The distance between spots 2 and 3 is 245.9 nm, between spots 3 and 1 is 318.3 nm, between spots 3 and 4 is 238 nm, and between spots 3 and 5 is 469 nm. The laser excitation is at 660 nm, and the laser filter blocks wavelengths shorter than ~665 nm. As a result, only a small portion of $WS_2$ PL is observed as shown in the green shaded area in figure 1c. Similarly, the red shaded area shows the PL of $MoS_2$.

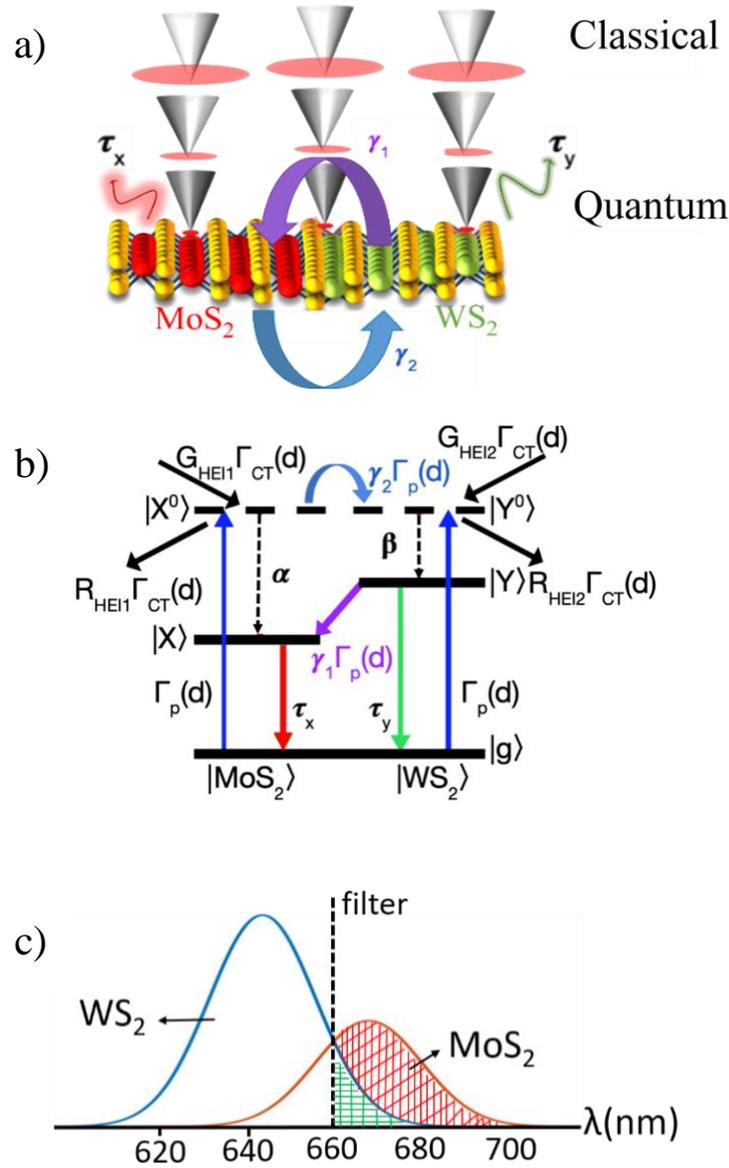

Figure 1. Lateral MoS$_2$-WS$_2$ heterostructure. (a) Sketch of the controlled tip-sample distance dependence measurements in the classical and quantum regimes. A 660 nm linearly polarized laser is on the apex of the Au-coated Ag plasmonic tip and the emitted PL signals are detected when the tip-sample distance is in the classical (d>0.36 nm) and in the quantum regime (d<0.36 nm). (b) Schematic state diagram at the junction of the 2D lateral heterojunction in tip-enhanced photoluminescence (TEPL) experiments. Hot electron injection (HEI) occurs from the plasmonic tip to the semiconductor. (c) Sketch of the PL peaks of WS$_2$ and MoS$_2$ and the laser. The green shaded area and red shaded area show the PL of WS$_2$ and MoS$_2$, respectively.

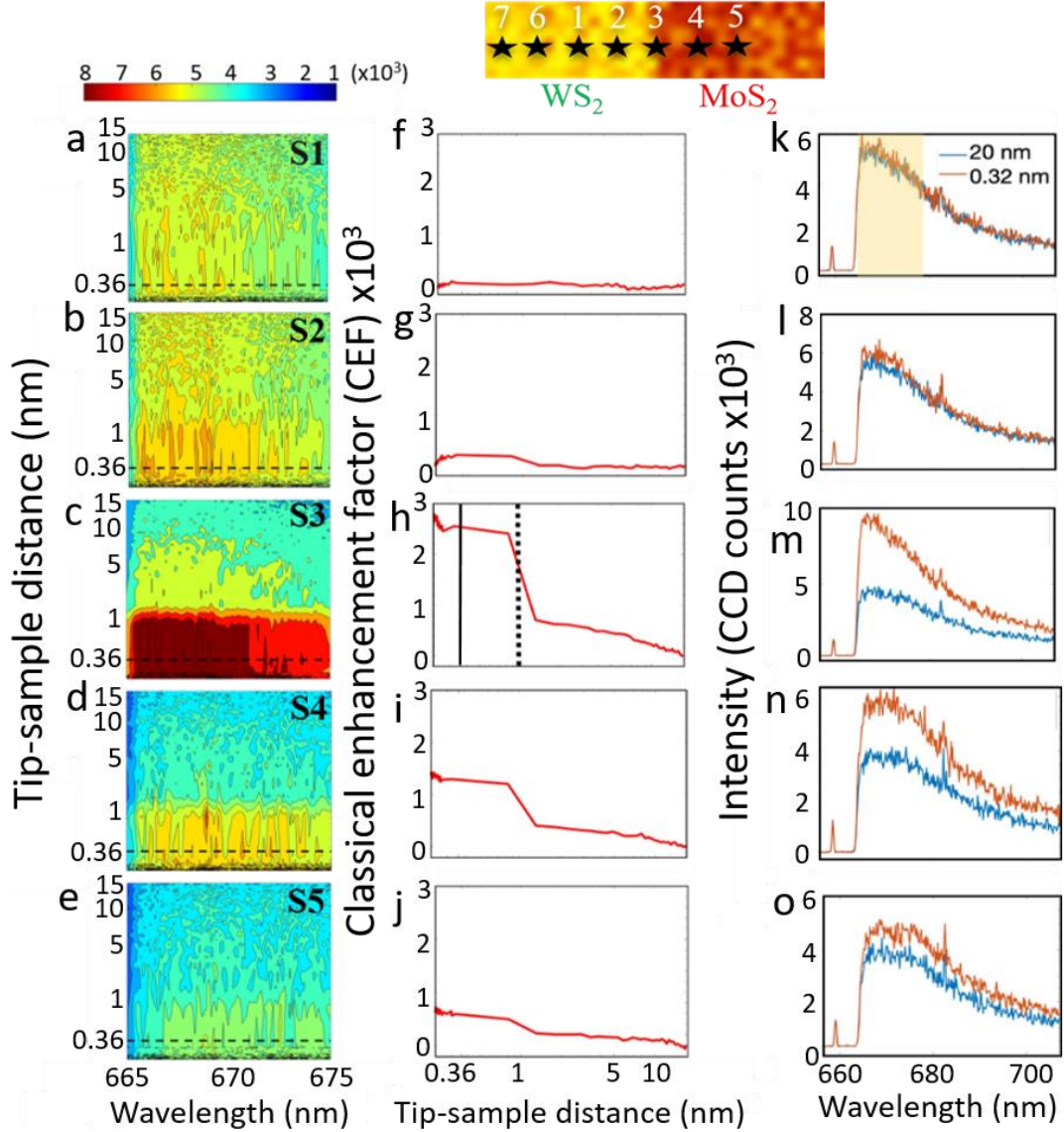

Figure 2. Tip-sample distance dependence PL measurements. 2D contour plots showing tip-sample distance measurements with 0.2 nm ≤ d ≤ 20 nm on 5 spots (a-e). At spot 3, a significant PL enhancement is observed at 1 nm tip-sample distance due to the hot electron enhancement mechanism (c, h, m). TEPL spectra at 5 spots (k-o).

The 2D PL intensity contour maps in figures 2a-2e show the picoscale controlled tip-sample distance measurements on five spots (S1 to S5). The picometer scale tip-sample distance approach has been previously developed[29] and utilized in recent experiments[29,33].

At spots 1 and 2, due to the laser filter, only a small emitted shoulder PL peak of $WS_2$ is observed as shown in figure 1c. This is the reason for the small change in PL intensity from the classical to the quantum regime. This can be verified by observing the spectra in figure 2k at d = 20 nm (classical) and d = 0.36 nm (quantum). Moving toward the heterojunction, a small enhancement is shown from 1 nm to 0.36 nm. However, at the heterojunction, as shown by spot 3, the enhancement in the PL intensity is observed from 20 nm to 1 nm and a much significant enhancement is observed from 1 nm to 0.36 nm. The enhancement at spot 3 is attributed to $WS_2$ because of the shape of the spectral distribution (Figures 1c and 2m). At spots 4 and 5, as shown in figures 2d, 2n, 2e and 2o, the enhancement of the PL intensity decreases from classical to quantum regime for the increasing lateral distance away from the junction. The enhancement factor as a function of the tip-sample distance has been calculated and plotted for all the 5 spots. In the classical regime, the furthest data point was the far field reference point. All the data was normalized to this data point and the enhancement factors (EFs) were calculated as previously described[29].

We define the chemical enhancement factor (CEF) and quantum enhancement factor (QEF) based on the classical (>1 nm) and quantum plasmonic ( <1 nm) distance regimes, respectively. In the quantum plasmonic regime, when the tip reaches at a distance of 0.36 nm, which is the vdW contact, directional hot electron tunneling takes place from the plasmonic tip to the sample. Previously, it has been shown, even in the absence of metal-metal contacts, quenching of $WSe_2$ PL has been observed at the heterojunction[33]. We show the similar effect of PL quenching of, $MoS_2$ on $SiO_2$/Si substrate at the heterojunction (figures 2 and 3).

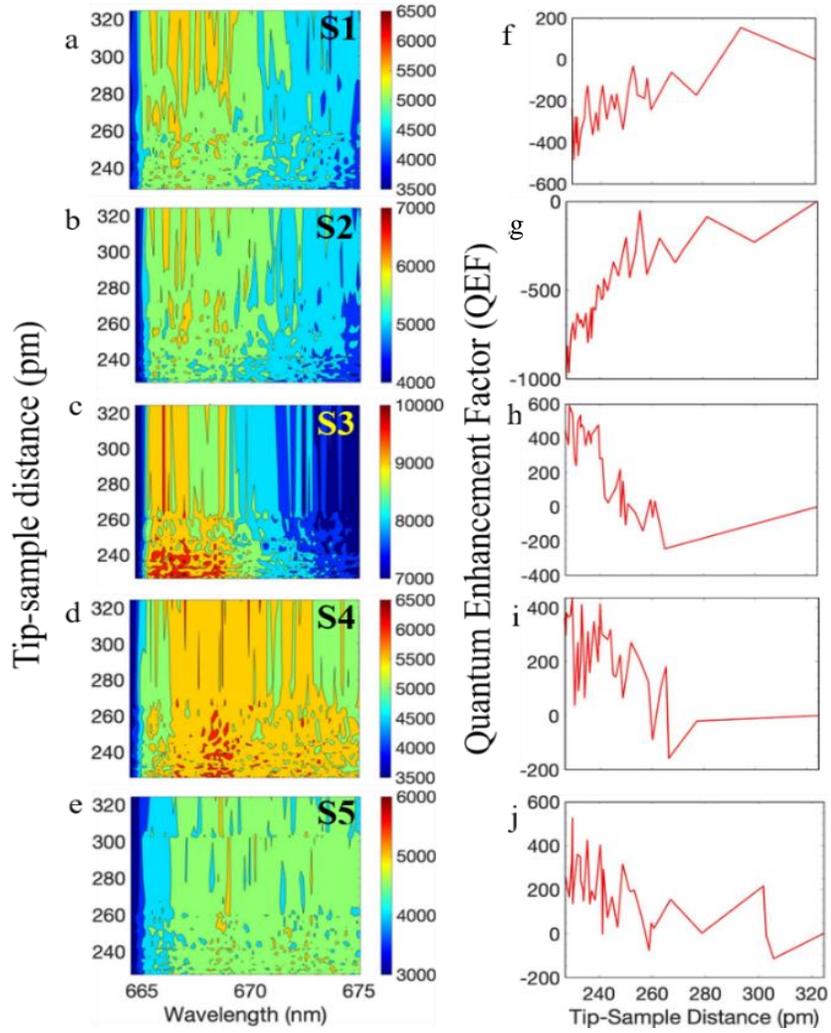

Figure 3. Tip-sample distance dependence measurements in the quantum plasmonic regime.

**Theoretical model**

A phenomenological rate equation model is used in order to further understand the interplay between hot electron injection and charge transfer during the experiment. Figure 1b shows the state diagram used to model these processes near the heterojunction. An initial population, N, of electrons is excited from the ground state, $|g\rangle$, to states $|X^0\rangle$ or $|Y^0\rangle$, by the near-field of the plasmonic AFM tip, which then decay to exciton states $|X\rangle$ or $|Y\rangle$. As a result, the population dynamics can be described by the rate equations:

$$\frac{dN_{X0}}{dt} = (G_{HEI_1} - R_{HEI_1}N_{X0})\Gamma_{CT}(d) - \alpha N_{X0} + \Gamma_p(d)(N_g - N_{X0}) - \gamma_2\Gamma_p(d), \quad (1)$$

$$\frac{dN_{Y0}}{dt} = (G_{HEI_2} - R_{HEI_2}N_{Y0})\Gamma_{CT}(d) - \beta N_{Y0} + \Gamma_p(d)(N_g - N_{Y0}) + \gamma_2\Gamma_p(d), \quad (2)$$

$$\frac{dN_X}{dt} = \alpha N_{X0} + \gamma_1\Gamma_p(d) - \frac{N_X}{\tau_X}, \quad (3)$$

$$\frac{dN_Y}{dt} = \beta N_{Y0} - \gamma_1\Gamma_p(d) - \frac{N_Y}{\tau_Y}, \quad (4)$$

$$\frac{dN_g}{dt} = -\Gamma_p(d)(N_g - N_{X0}) - \Gamma_p(d)(N_g - N_{Y0}) + \frac{N_X}{\tau_X} + \frac{N_Y}{\tau_Y}. \quad (5)$$

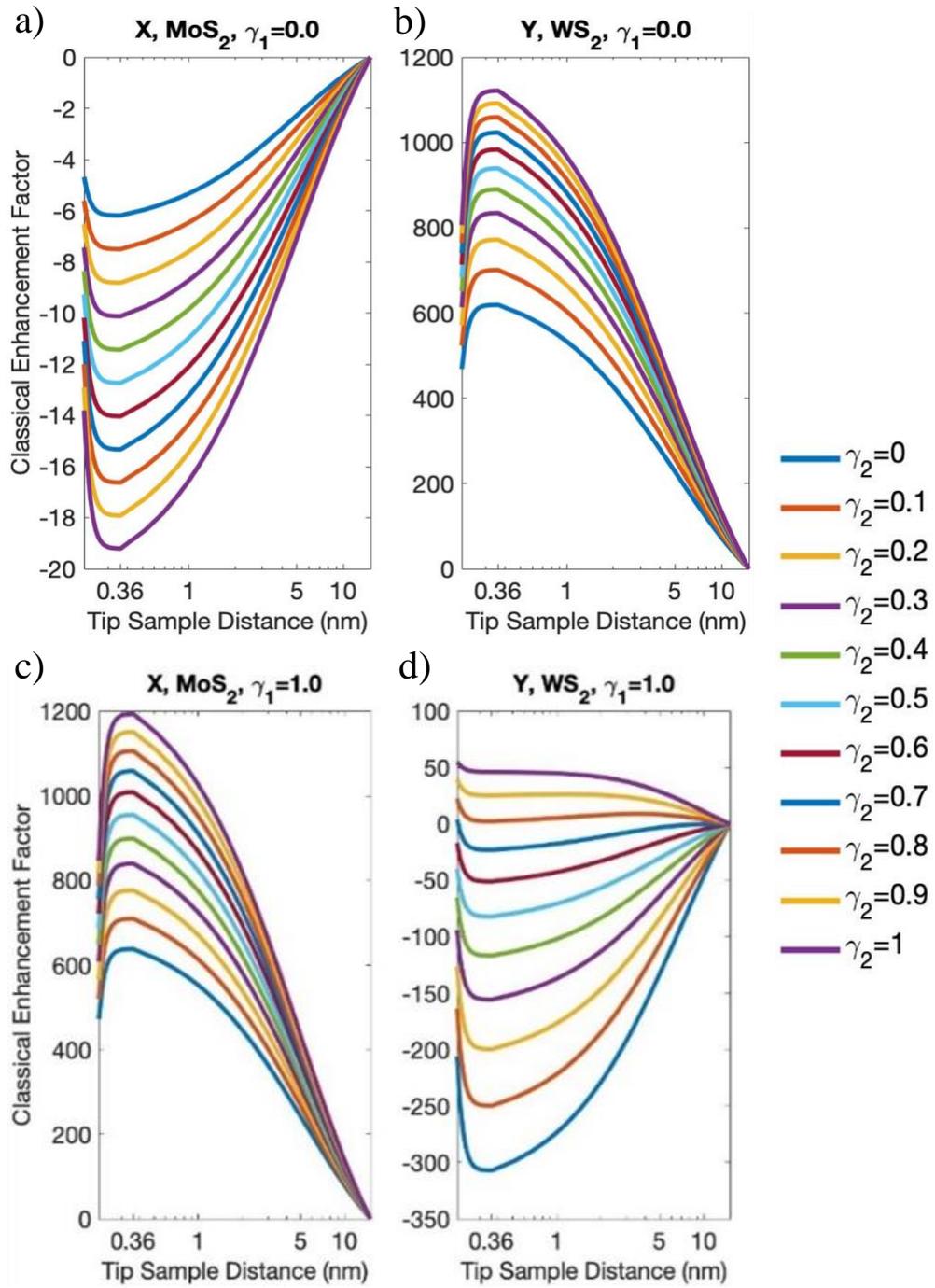

Figure 4. Simulated tip-sample distance dependent classical enhancement factors of the PL at the heterojunction. The charge transfer coefficient, $\gamma_1 = 0$ (a,b) and $\gamma_1 = 1$ (c,d) for several values of the charge transfer coefficient, $\gamma_2$.

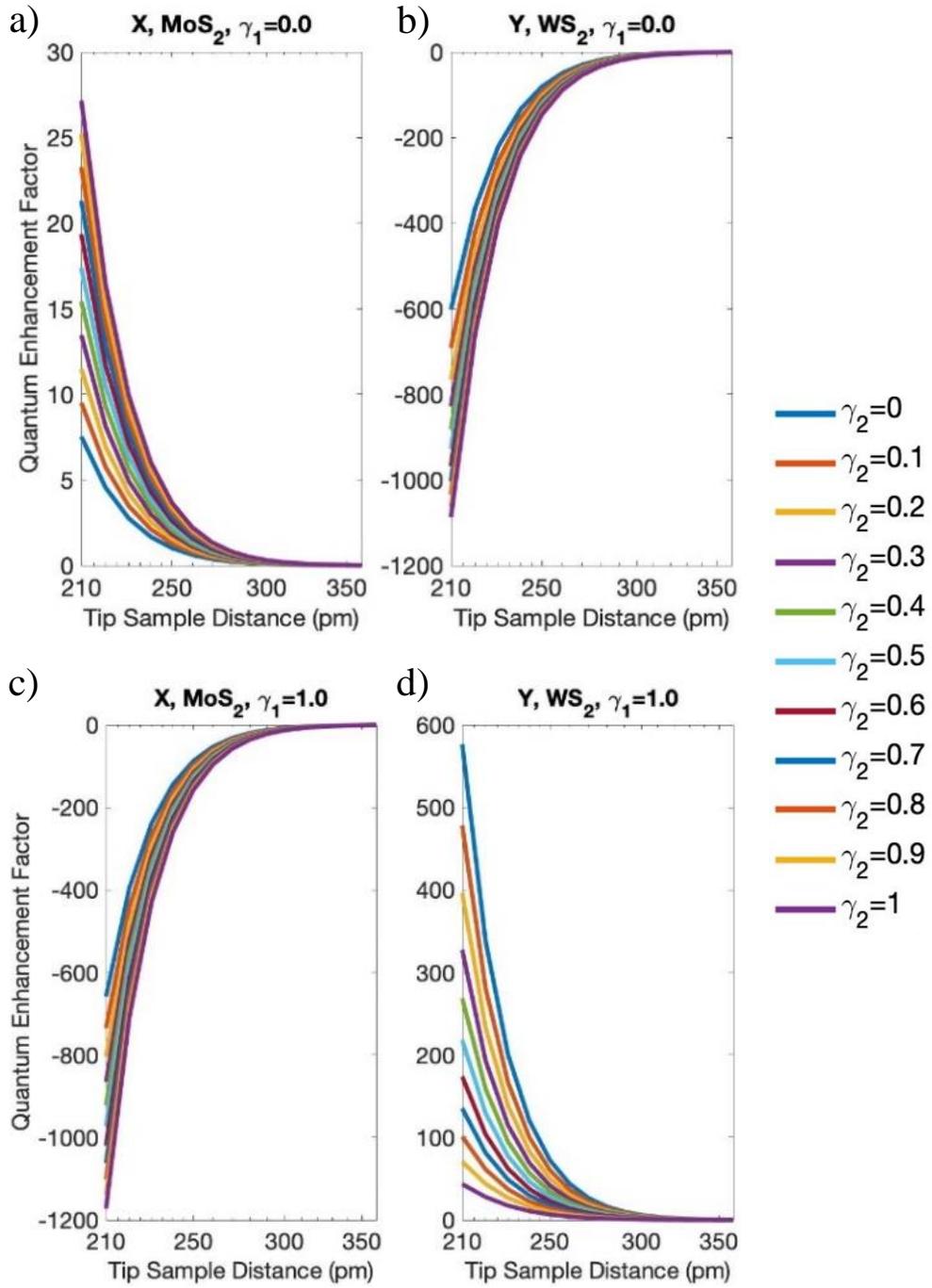

Figure 5. Simulated tip-sample distance dependent quantum enhancement factors (QEF) of the PL at the heterojunction. The charge transfer coefficient, $\gamma_1$ = 0 (a,b) and $\gamma_1$ = 1 (c,d) for several values of $\gamma_2$.

The plasmonic near field is used for the excitation (blue arrows) with the rate[30],

$$\Gamma_p(d) = \begin{cases} 1 - e^{-((d-c)/d_p)} & \text{for } c < d < 0.36 \text{ nm} \\ \dfrac{1}{A_p}\left(1 - \dfrac{B}{(R+d-c)^3}\right)^2 & \text{for } d > 0.36 \text{ nm} \end{cases}$$

where $d_p = 0.02$ nm is the average quantum coupling distance, $c = 0.17$ nm is the ohmic contact distance, $R = 25$ nm is the radius of the tip apex, $A_p$ ensures continuity, and $B = 5028$ includes the polarizability of the tip [30].

Hot electron tunnelling effects (black arrows) are modelled by[33],

$$\Gamma_{CT}(d) = \begin{cases} A_{CT} e^{-((d-c)/d_{CT})} & \text{for } d < 0.36 \text{ nm} \\ 0 & \text{for } d > 0.36 \text{ nm} \end{cases}$$

where $d_{CT} = 0.02$ nm is the average tunnelling distance, and A = 1 is a normalization parameter[33]. Figures 4 and 5 show the simulated PL classical and quantum enhancement factors, respectively, for $\gamma_1 = 0$ (a,b) and $\gamma_1 = 1$ (c,d) for several values of the charge transfer coefficient, $\gamma_2$. The simulations show agreement with the experiments for a range of parameters.

**Acknowledgements**

We thank the Prof. Prasana Sahoo for help with sample preparation. D.V.V. acknowledges the support by the National Science Foundation (NSF; Grant No. CHE-1609608).